\documentclass{nature}
\usepackage{deluxetable}
\usepackage{psfig}

\def\farcs{\hbox{$.\!\!^{\prime\prime}$}}
\def\la{\mathrel{\hbox{\rlap{\hbox{\lower4pt\hbox{$\sim$}}}\hbox{$<$}}}}
\def\ga{\mathrel{\hbox{\rlap{\hbox{\lower4pt\hbox{$\sim$}}}\hbox{$>$}}}}

\title{Long $\gamma$-ray bursts and core-collapse supernovae have different environments}

\author{A.S.~Fruchter$^{1}$, A.J.~Levan$^{1,2,3}$, L.~Strolger$^{1,4}$, P.M.~Vreeswijk$^5$, S.E.~Thorsett$^6$, D.~Bersier$^{1,7}$, I.~Burud$^{1,8}$, J.M.~Castro~Cer\'{o}n$^{1,9}$, 
A.J.~Castro-Tirado$^{10}$, C.~Conselice$^{11,12}$, T.~Dahlen$^{13}$, H.C.~Ferguson$^1$, J.P.U.~Fynbo$^9$, P.M.~Garnavich$^{14}$, R.A.~Gibbons$^{1,15}$, J.~Gorosabel$^{1,10}$, T.R.~Gull$^{16}$, J.~Hjorth$^9$, S.T.~Holland$^{17}$, C.~Kouveliotou$^{18}$, Z.~Levay$^1$, M. Livio$^1$, M.R.~Metzger$^{19}$,
P.E.~Nugent$^{20}$, L.~Petro$^1$, E.~Pian$^{21}$, J.E.~Rhoads$^1$, A.G.~Riess$^1$, K.C.~Sahu$^1$, A.~Smette$^5$, N.R.~Tanvir$^{3}$, R.A.M.J.~Wijers$^{22}$, S.E.~Woosley$^6$}

\begin{document}
\maketitle

\begin{affiliations}
 \item Space Telescope Science Institute, 3700 San Martin Dr., Baltimore, MD 21218, USA
 \item Department of Physics and Astronomy, University of Leicester, University Road, Leicester, LE1 7RH, UK
  \item Centre for Astrophysics Research, University of Hertfordshire, College Lane, Hatfield, AL10 9AB, UK
 \item Physics \& Astronomy, TCCW 246, Western Kentucky University, 1 Big Red Way, Bowling Green, KY 42101, USA
 \item European Southern Observatory, Alonso de C\'ordova 3107, Casilla 19001, Santiago, Chile
 \item Dept of Astronomy \& Astrophysics, University of California, 1156 High St, Santa Cruz, CA 95064, USA
 \item Astrophysics Research Institute, Liverpool John Moores University, Twelve Quays House, Egerton Wharf, Birkenhead, CH41 1LD  
 \item Norwegian Meteorological Institute, P.O. Box 43, Blindern, N-0313 Oslo, Norway
 \item Dark Cosmology Centre, Niels Bohr Institute, University of Copenhagen, DK-2100 Copenhagen, Denmark
 \item Instituto de Astrof\'{\i}sica de Andaluc\'{\i}a (CSIC), Camino Bajo de Hu\'etor, 50, 18008 Granada, Spain.
 \item California Institute of Technology, Mail Code 105-24, Pasadena, CA 91125, USA
 \item School of Physics and Astronomy, University of Nottingham, University Park, United Kingdom, NG7 2RD
 \item Department of Physics, Stockholm University, SE-106 91 Stockholm, Sweden
 \item Physics Department, University of Notre Dame, 225 Nieuwland Hall, Notre Dame, IN 46556, USA
 \item Vanderbilt University, Dept. of Physics and Astronomy, 6301 Stevenson Center, Nashville, TN 37235, USA
 \item Code 667 Extraterrestial Planets and Stellar Astrophysics, Exploration of the Universe Division, Goddard Space Flight Center, Greenbelt, MD 20771, USA
 \item  Code 660.1, NASA's GSFC, Greenbelt, MD 20771, USA
 \item NASA/Marshall Space Flight Center, VP-62, National Space Science \& Technology Center, 320 Sparkman Drive, Huntsville, AL 35805, USA
 \item Renaissance Technologies Corporation, 600 Route 25A, East Setauket, New York 11733
 \item Lawrence Berkeley National Laboratory, M.S. 50F-1650, 1 Cyclotron Road, Berkeley, CA 94720, USA
 \item INAF, Osservatorio Astronomico di Trieste, Via G.B. Tiepolo 11, I-34131 Trieste, Italy
 \item  Astronomical Institute `Anton Pannekoek', University of Amsterdam, Kruislaan 403, NL-1098 SJ Amsterdam, The Netherlands 
\end{affiliations}

\clearpage

{\it This paper has been accepted for publication in {\rm Nature}.   In accordance with that journal's editorial policy we ask that you do not discuss this work with the Press until it appears in {\rm Nature} either online or in print.  Thank you.} \\

\begin{abstract}

When massive stars exhaust their fuel they collapse and often produce the extraordinarily
bright explosions known as core-collapse supernovae.   
On occasion, this stellar collapse also
powers an even more brilliant relativistic explosion known as a
long-duration $\gamma$-ray burst.
One would then expect that long $\gamma$-ray bursts and core-collapse supernovae
should be found in similar galactic environments.  Here we show that this
expectation is wrong.   We find that
the long $\gamma$-ray bursts are far more concentrated on the
very brightest
regions of their host galaxies than are the core-collapse supernovae.
Furthermore, the host galaxies of the long $\gamma$-ray bursts are significantly fainter and more irregular than the
hosts of the core-collapse supernovae. Together these
results suggest that long-duration $\gamma$-ray bursts are associated with the most massive stars and
may be restricted to galaxies of limited chemical evolution.   Our results directly imply that long $\gamma$-ray bursts are relatively rare in galaxies such as our own Milky Way.

\end{abstract}

It is an irony of astrophysics that stellar birth is most spectacularly marked
by the deaths of massive stars. Massive stars burn brighter and hotter than
smaller stars, and exhaust their fuel far more rapidly. Therefore a
region of star formation filled with low mass stars still early in their lives, and in
some cases still forming, may also host massive stars already collapsing and
producing supernovae. Indeed, with the exception of the now famous Type Ia
supernovae , which have been so successfully used for cosmological studies\cite{rfc+98,pag+99}
and which
are thought to be formed by the uncontrolled nuclear burning of stellar remnants
comparable in mass to the sun\cite{bly+05}, all supernovae are thought to be produced by the
collapse of massive stars. The collapse of the very most massive stars (tens of
solar masses) is thought to leave behind either black holes or neutron stars,
depending largely on the state of chemical evolution of the material that formed
the star, while the demise of stars between approximately 8 and 20 solar
masses produces only neutron stars\cite{hfw+03}.

Gamma-ray bursts (GRBs), like supernovae, are a heterogeneous population. GRBs can be divided into two
classes: short, hard bursts, which last between milliseconds and about two
seconds and have hard high-energy spectra, and long, soft bursts, which
last between two and tens of seconds, and have softer high-energy spectra\cite{kmf+93}.
Only very recently have a few of the short bursts been well localized, and 
initial studies of their apparent hosts suggest that these bursts may be formed 
by the binary merger of stellar remnants\cite{gso+05,pbc+05}.
In contrast, the afterglows of over eighty long GRBs (LGRBs) have been detected
in the optical and/or radio. And as a result of these detections, it has become
clear that LGRBs, like core-collapse supernovae, are related to the deaths
of young, massive stars.
It is these objects, born of the deaths of massive
stars, that we study here.  

LGRBs are generally found in
extremely blue host galaxies\cite{ftm+99,sfc+01,ldm+03,chg04} which exhibit strong emission lines\cite{bdkf98,vfk+01}
suggesting a significant abundance of young, very massive stars. Furthermore
while the light curves of the optical transients (OTs) associated with LGRBs are
often dominated by radiation from the relativistic outflow of the GRB,
numerous LGRBs have shown late-time ``bumps'' in their light-curves consistent with
the presence of an underlying SN\cite{bkd+99,gtv+00,lnf+05}.  In several cases spectroscopic evidence
has provided confirmation of the light of a SN superposed on the OT\cite{hsm+03,smg+03,dmb+03,mtc+04}. Indeed,
given the large variations in the brightnesses of OTs and supernovae, and the limited
observations on some GRBs, it seems plausible that {\it all} LGRBs
have an underlying SN\cite{zkh04}. Furthermore, while the energy released in a LGRB often appears to
the observer to be orders of magnitude larger than that of a supernovae, there is
now good evidence suggesting that most LGRBs are highly collimated and often
illuminate only a few percent of the sky\cite{pk01,fks+01}. When one takes this into account, the
energy released in LGRBs more closely resembles that of energetic supernovae.    
However, not
all core-collapse supernovae may be candidates for the production of GRBs.    The supernovae with good spectroscopic identifications so far associated with GRBs have been Type
Ic -- that is cc supernovae which show no evidence of hydrogen or helium in their spectra. (Type Ib supernovae,
which are often studied together with Type Ic, have spectra which are also largely devoid of hydrogen lines but show strong helium features.)    A star may therefore need to lose its outer envelope if a GRB is to be able to
burn its way through the stellar atmosphere\cite{mwh01}.  Studies which have compared the locations Type Ib/c supernovae with the more numerous Type II supernovae (cc supernovae showing hydrogen lines) in local galaxies so far show no differences in either the type of host or the placement of the explosion on the host \cite{vhf96,vdblf05}.    This result led Ref.~\citen{vhf96} to argue cc supernovae all come from the
same mass range of progenitor stars, but that Type Ib/c supernovae may have had their envelopes
stripped by interaction with a binary stellar companion.
Whether Type Ic supernovae come from single stars, or binary stars, or both, it  is very likely
that only small fraction of these supernovae produce GRBs\cite{pmn+04}.  

Given the common massive stellar origins of cc supernovae and LGRBs, one might expect that their hosts and 
local environments might be quite similar.   It has long been argued that cc supernovae should track the blue
light in the universe (the light from massive stars is blue), both in their distribution among galaxies and within their host galaxies themselves.   One would expect similar behavior from LGRBs, and indeed
rough evidence for such a correlation has been reported\cite{bkd02}.    Here we use the high resolution available
from Hubble Space Telescope ({\it HST})
 images, and an analytical technique developed by us that is independent of galaxy morphology,
to study the correlation between these objects and the light of their hosts.    We also compare
the sizes, morphologies and brightnesses of the LGRB hosts with those of the supernovae.  Our results reveal
surprising and substantial differences between the birth places of these cosmic explosions.    We find
that while cc supernovae trace the blue-light of their hosts, GRBs are far more concentrated on the brightest regions of their hosts.  Furthermore, while the hosts of cc supernovae are approximately equally divided between spiral and irregular galaxies, the overwhelming majority of GRBs are on irregulars, even when we restrict the GRB sample to the same redshift range as the SN sample.  We argue that these results may be best understood if GRBs are formed from the collapse of extremely massive, low-metallicity stars.

\section{The Sample}

Over forty LGRBs have been observed with {\it HST} at various times after outburst.  {\it HST} is unique in its capability to easily resolve
the distant hosts of these objects.   Shown in Figure~1 is a mosaic of {\it HST} images of the hosts of forty-two bursts.   These are all LGRBs with public data  which had an afterglow detected with better than three-sigma significance and a position sufficiently well localized to determine a host
galaxy.    A list of all the GRBs used in this work can be found in Tables 1---3 of the Supplementary Material.  
 

The supernovae discussed in this {\it Article} were all discovered as part of the  {\it Hubble} Higher z Supernova Search\cite{rst+04,srd+04}, which was done in cooperation with the {\it HST} GOODS survey\cite{gfk+04}.  The GOODS survey observed two $\sim150$ sq.~arcminute patches of sky five times each in epochs separated by forty-five 
days.    Supernovae were identified by image subtraction.   In this
paper we discuss only the cc supernovae identified in this survey.  A list of the supernovae used is presented in Table 4 of the Supplementary Material,
and images of the supernovae hosts can be seen in Figure 2.

\section{Positions of GRBs and supernovae on their Hosts}

If LGRBs do in fact trace massive star formation, then in the absence of strong
extinction we should find a close correlation between
their position on their host galaxies and the blue light of those galaxies.
However, many of the GRB hosts and quite a few of the supernovae hosts are irregular galaxies made up of more than one bright component.  As a result  the common astronomical procedure of identifying the centroid of the galaxy's light, and then determining the distance of the object in question from the centroid is not particularly appropriate for these galaxies -- the centroid of
light may in fact lie on a rather faint region of the host (examine GRBs 000926 and 020903 in Figure 1 for excellent
illustrations of this effect).   We therefore have developed a method which
is independent of galaxy morphology.  We sort all of the pixels
of the host galaxy image from faintest to brightest and ask what fraction of the total light of the host is contained in pixels fainter than or equal to the pixel containing the explosion.    
If the explosions track the distribution of light, then the fraction determined by this method should be uniformly distributed between zero and one.  (A detailed exposition of this method can be found
in the supplementary materials).

As can be seen in Figure 3, the cc supernovae do track the light of their hosts as well as could be expected given their small number statistics.    A Kolmogornov-Smirnov (KS) test finds that the distribution of the supernovae is indistinguishable from the distribution of the underlying light.    The situation is clearly different for LGRBs.
As can be seen in Figure 3, the GRBs do not simply trace the blue light of the hosts, rather they are far more concentrated on 
the peaks of light in the hosts than the light itself.  A KS test rejects the hypothesis that GRBs are distributed as the light of their hosts with a probability greater than 99.98\%.   Furthermore, this result
is robust:  it shows no dependence on GRB host size or magnitude.  And in spite of the relatively
small number of SN hosts on which a comparison can be made, the two populations are found
by the KS test to be drawn from different distributions with $\sim 99 \%$ certainty.   In
the next section of this paper we show that the surprising differences in the locations of these objects on the underlying
light of their hosts may be due not only to the nature of their progenitor stars but also that of their hosts. 

\section{A Comparison of the Host Populations}

An examination of the mosaics of the GRB and SN hosts (Figures 1 and 2)  immediately shows a remarkable contrast -- only one GRB host in this set of 42 galaxies is a grand-design spiral, while nearly half of the SN hosts are grand-design spirals.  One might wonder if this effect is due to a difference in redshift distribution -- the cc supernovae discovered by the GOODS collaboration all lie at $z<1.2$, while LGRBs can be found at much larger redshifts where grand-design spirals are rare to non-existent.  Yet if we restrict the GRB population to $z<1.2$ (and thus produce a population with a nearly identical mean and standard deviation in redshift space compared to the GOODS cc supernovae), the situation remains essentially unchanged:  only one out of the eighteen GRB hosts is a grand-design spiral.  (For a detailed comparison of GRB hosts to field galaxies, rather than the SN selected galaxies shown here, see Ref.~\citen{cvf+05}).

Were the difference in spiral fraction the only indication of a difference in the host populations, we could not rule out random chance -- given the small number statistics both populations are barely consistent with each other and a spiral fraction of $\sim 25\%$.  However,  the host populations differ strongly in ways other than morphology.   

In Figure~4 we compare the 80\% light radius ($r_{80}$) and
 absolute magnitude distributions of the GRB and
supernovae hosts.   Included in the comparison are all LGRBs with known redshifts $z < 1.2$  at the time
of submission and the
16 cc supernovae of GOODS  with spectroscopic or photometric redshifts (See the Supplementary Tables
for a complete list of the GRBs, supernovae and associated parameters used in this study).  
The small minority of GRB
hosts in this redshift range without {\it HST}  imaging are compared only in absolute magnitude and not in size.   The
absolute magnitudes have been derived from the
observed photometry using a cosmology of $\Omega_m=0.27,
\Lambda=0.73$, and $H_0=71 {\rm km s^{-1} Mpc^{-1}}$, and the magnitudes
have been corrected for foreground Galactic extinction\cite{sfd98}.  For a technical discussion of the determination of the magnitude and size of individual
objects, please see the Supplementary Materials.  

As can be readily seen the two host populations differ substantially both in their intrinsic magnitudes
and sizes.  The GRB hosts are fainter and smaller than the SN hosts.  Indeed KS tests
reject the hypothesis that these two populations are drawn from the same population  
with greater than 98.6\% and 99.7\% certainty for the magnitude and size distributions, respectively.

\section{Discussion}
  
Although the evidence is now overwhelming that both cc supernovae and LGRBs are formed by the collapse of 
massive stars, our observations show that the distribution of LGRBs and cc supernovae on
their hosts, and the nature of their hosts themselves are substantially different.  How 
then can this be?
We propose here that these surprising findings are the result of the dependence of
the probability of GRB formation on the state of the chemical evolution of massive stars
in a galaxy.

Even before the association of LGRBs with massive stars had been established,   a number of
theorists had suggested that these objects could be formed by the collapse of  massive
stars, which would leave behind rapidly spinning black holes.   An accretion disk about
the black hole would power the GRB jet.   These models, sometimes referred to as 
``hypernovae" or ``collapsar" models implicitly require very massive stars, since only
stars greater than about $18$ solar masses form black holes.   But in fact it was widely
suspected that even more massive stars would be required -- if only to provide the required
large energies, and to limit the numbers of supernovae progressing to GRBs.

We conclude that LGRBs do indeed form from the very most massive
stars and this is the reason that they are even more concentrated on the 
blue light of their hosts than the light itself.  The most massive stars (O stars) are frequently found in 
large associations.  These associations can be extremely
bright, and can indeed provide the peak of the light of a galaxy -- particularly
if that galaxy is a faint, blue irregular, as are the GRB hosts in general.   
Indeed, a connection of LGRBs with O-stars (and perhaps Wolf-Rayet stars)
is a natural one -- given the strong emission lines (including Ne [III]) seen in many of these hosts\cite{bdkf98,vfk+01} and the evidence for possible strong winds off of the progenitors
of the GRBs seen in absorption in some LGRB spectra\cite{mhc+03,kgr+04}.

However, O stars are found in galaxies of all sizes.  Indeed, studies of the Magellanic clouds
suggest the initial distribution of masses of stars at formation in these dwarf galaxies is essentially
identical to that in our much larger spiral, the Milky Way\cite{wk05b}.   Therefore, a difference in
the initial mass function of stars
 is unlikely to be responsible for the differences between the hosts.   We propose
that the fundamental differences between the LGRB and SN host populations is
not their size or luminosity, but rather their metallicity, or chemical evolution.
Some evidence of this already exists.   The hosts of seven LGRBs (GRBs 980425 (P.~M.~Vreeswijk, personal
communication),
990712\cite{vfk+01}, 020903\cite{bfs+05}, 030323\cite{vel+04}, 030329\cite{hsm+03}, 031203\cite{pbc+04}
and 050730\cite{cpb+05}) 
 have
measurements of or  limits on their metallicity, and in all cases the metallicity
is less than one-third solar.   The small size and low luminosity of the
GRB hosts is then a result of the well known correlation between
galaxy mass and metallicity (see Ref.~\citen{kk04} and references therein).   

But why do LGRBs choose low-metallicity
galaxies?  This may be a direct result of the evolution of the most
massive stars.    It has recently been proposed that
metal rich stars with masses of tens of solar masses
have such large winds off their surfaces (due to the photon pressure on
their metal rich atmospheres) that they lose most of their mass before
they collapse and produce supernovae\cite{hfw+03}.  As a result they leave behind
neutron stars, not the black holes necessary for LGRB formation.  Ironically,
stars of 15-30 solar masses may still form black holes, as they do
not possess radiation pressure sufficient to drive off their outer
envelopes.
Direct evidence for this scenario comes from recent work showing that the Galactic
soft gamma-ray repeater,
SGR 1820-06, is in a cluster of extremely young stars of 
which the most massive have only started to collapse\cite{fng+05} -- yet, the progenitor
of SGR 1820-06 collapsed to a neutron star, not a black hole.  Recent observations
of winds from very massive (Wolf-Rayet) stars provide
further support for this scenario:  outlfows
from the low-metallicity stars in the LMC are substantially smaller than
those seen from more metal-rich Galactic stars\cite{ch06}.   The 
possible importance of metallicity in LGRB formation has therefore
not escaped the notice of theorists\cite{wh05,yl05}.

A preference for low-metallicity may also explain one of the most puzzling results of GRB
host studies.  None of the LGRB hosts is a red, sub-millimeter bright
galaxy.   These highly dust-enshrouded galaxies at redshifts of $\sim 1-3$
are believed to be the site of a large fraction of the star formation in the
distant universe\cite{cbs+05}.   And while some LGRB hosts do show sub-mm emission, 
none has the red colors characteristic of the majority of this population.
However, it is likely that these red dusty galaxies have substantial
metallicities at all redshifts.  
The low-metallicity of hosts may also help explain the fact that a substantial
fraction of high-redshift LGRB hosts display strong Lyman-alpha emission\cite{fjm+03}.

 All well classified supernovae associated with LGRBs are Types  Ic,  presumably because the presence of
a hydrogen envelope about the collapsing core can block the emergence of
a GRB jet\cite{mwh01}.    Thus only those supernovae whose progenitors have lost some, but
not too much mass, appear to be candidates for the formation of a GRB.
Given the large numbers of Type Ic supernovae in comparison to the estimated
numbers of LGRBs however, it is likely that only a small fraction of
Ic supernovae produce LGRBs.  Indeed, even the number of unusually
energetic Type Ib/Ic supernovae appears to dwarf the LGRB population\cite{snk+06}.
 Another process, perhaps
the spin-up of the progenitor in a binary\cite{pmn+04}, may decide which Type
Ic supernovae produce LGRBs.    Interestingly, it was the similar distribution of
supernovae on their hosts, and particularly the fact that Type Ib/Ic were no more correlated
than Type II supernovae with the UV bright regions of their hosts,  that led Ref~\citen{vhf96} to the conclusion that Type Ib/Ic 
form from binaries.    LGRBs clearly track light differently than the general 
Type Ic population.   However the samples used by Refs~\citen{vhf96,vdblf05} were from 
supernovae largely discovered on nearby massive galaxies -- dwarf irregular hosts are underrepresented
in these samples.  It will be particularly interesting to see whether large unbiased SN surveys
presently underway produce similar locations for their supernovae.

We do not know, however, what separates the small fraction of low-metallicity Type Ic supernovae
which turn into LGRBs from the rest of the population.  Potentially, the answer is the amount
of angular momentum available in the core to form the jet.   In this case, the preference for
low-metallicity may indicate that single star evolution dominates over binary interaction
in forming LGRBs.   Deep, high spectral resolution studies of LGRB afterglows may provide
insight here, by allowing a studies of the winds off of the progenitor and any binary companion.

Only a small fraction of LGRBs are found in spiral galaxies, even in LGRBs with  redshifts
$z < 1$ where spirals are much more common.    However, the local metallicity in spirals
is known to be anti-correlated with distance from the center of the galaxy.  Thus one might
expect LGRBs in spirals to violate the trend we have seen for the general LGRB population
and avoid the bright central regions of their hosts.   The present number of LGRBs
known in spirals is still too small to test this prediction.  But a sample size a few times larger
should begin to allow such a test.    Additionally, a survey of the metallicity of the hosts of the
GOODS supernovae should find a higher average metallicity than that seen in GRB hosts.
Finally, if low-metallicity is indeed the primary
variable in determining whether LGRBs are produced,
then as we observe higher redshifts, where metallicities are lower
than in most local galaxies, LGRBs should be more
uniformly distributed among star-forming galaxies.  Indeed, some evidence of this may already
be present in the data\cite{cvf+05}.   LGRBs, however, are potentially visible to redshifts as high as $z \sim 10$.   At significant redshifts, where the metallicities of even
relatively large galaxies are expected to be low, we may find that LGRBs do
become nearly unbiased tracers of star formation.

\eject

Figure I:  A mosaic of GRB host galaxies imaged by HST.   Each individual image corresponds
to a square region on the sky $3\farcs75$ on a side. 
  These images were taken with the Space Telescope Imaging Spectrograph
(STIS),  Wide-Field and Planetary Camera 2 (WFPC2) and the Advanced
Camera for Surveys (ACS) on {\it HST}. 
 In cases where the location of the GRB on the host
 is known to better than $0\farcs15$, the position of GRB is shown by a green mark.   If the positional error is smaller the the point spread function of the image ($0\farcs07$ for STIS and ACS, $0\farcs13$
 for WFPC2) the position is marked by a cross-hair, otherwise the positional error is indicated
 by a circle.   The STIS images were all taken in white light (no filter)
and in most cases the WFPC2 and ACS image are in the F606W filter (though in a few cases
where images in this filter were not available we have used images in F555W or 
F775W).   The STIS and F606W images can be thought of as  broad
"V" or visual images, and are, for galaxies exhibiting typical colors of GRB hosts,
 the single most sensitive settings for these cameras.   F555W is close to the ground-based Johnson V-band, and F775W corresponds to the ground-based Johnson I-band.
Due to the redshifts of the hosts, these images generally correspond
to blue or ultra-violet images of the hosts in their rest frame, and thus detect light largely produced by
the massive stars in the hosts.

Figure 2:  A mosaic of cc SN host galaxies imaged with {\it HST} as part of the GOODS program.
 Each image in the mosaic has a width of $7\farcs5$ on the sky, and thus two times the field-of-view
 of each image in the GRB mosaic. The position
 of each SN on its host galaxy is marked.  In all cases, these positions are known to sub-pixel accuracy.  supernovae in the GOODS sample were identified by [\citen{srd+04}] as either Type Ia or cc supernovae based on their colors, luminosities and light curves, as data allowed (a SN going off near the beginning or end of one of the multi-epoch observing runs  would have much less data, and sometimes poor color information).   Thus bright Type Ib and Ic supernovae, which have colors and luminosities similar to Type Ia supernovae, would have likely been classified as Type Ia (unless a grism spectrum was taken -- however only a small fraction of objects were observed spectroscopically).   On the other hand fainter Type Ib and Ic supernovae ($M_B \ga -18$) could in principle be identified from photometric
data;    however, in practice the data were rarely sufficient for a clear separation from other cc supernovae.   Based on
surveys of nearby galaxies, one might expect  approximately 20\% of the cc supernovae to be
Type Ib or Ic\cite{vdbt91,mdvp+05}.

Figure 3:  The locations of the explosions in comparison to the host light.   For
each object an arrow 
indicates the fraction of total host light in pixels fainter than or equal to the light
in the pixel at the location of the transient.  The cumulative fraction of GRBs or supernovae found at
a given fraction of the total light is shown as a histogram.   
 The blue arrows and histogram
correspond to the GRBs and the red arrows and histogram correspond to the supernovae.
Were the GRBs and supernovae
to track the light identically, their histograms would follow the diagonal line.  While the supernovae
positions do follow the light within the statistical error,  the GRBs are far more concentrated on the brightest regions of their hosts.   Thus while the probability of a SN exploding in a particular pixel is
roughly proportional to the surface brightness of the galaxy at that pixel, the probability of a GRB a given location effectively goes as a higher power of the local surface brightness.

Figure 4:  A comparison of the absolute magnitude and size  distributions of the GRB and SN hosts.  In the main panel,
the cc SN hosts are
represented as red squares and the LGRB hosts as blue circles.
The absolute magnitudes of the hosts are shown on
 the x-axis and the lengths of the
semi-major axes of the hosts  on the y-axis.   The plot is then projected onto the two side panels 
where a histogram is displayed for each host population in each of the dimensions - absolute magnitude and semi-major axis.   Shown as blue arrows are the absolute magnitudes
of GRB hosts with $z < 1.2$ that have been detected from the ground but have not yet been observed
by HST.   These hosts are only included in the absolute magnitude histogram.  The
hosts of GRBs are both smaller and fainter than those of supernovae.

\begin{addendum}
 \item  Support for this research was provided by NASA through a grant from the Space Telescope Science Institute, which is operated by the Association of Universities for Research in Astronomy, Inc.  Observations analyzed in this work were taken by the NASA/ESA Hubble Space Telescope
 under programs: 7785, 7863, 7966, 8189, 8588, 9074 and 9405 (PI: Fruchter);  7964, 8688, 9180, and 10135 (PI S.~R.~Kulkarni); 8640 (PI: S.~T.~Holland).   
The authors wish to thank Nino Panagia, Nolan Walborn and Alicia Soderberg
for informative conversations.  We also thank Alex Filippenko and collaborators for early-time images
of GRB~980326, and Josh Bloom and collaborators for making their early observations of
GRB~020322 public.
 \item[Competing Interests] The authors declare that they have no
competing financial interests.
 \item[Correspondence] Correspondence and requests for materials
should be addressed to Andrew Fruchter~(email: fruchter@stsci.edu).
\end{addendum}

\begin{figure*}[h]
\begin{center}
\psfig{figure=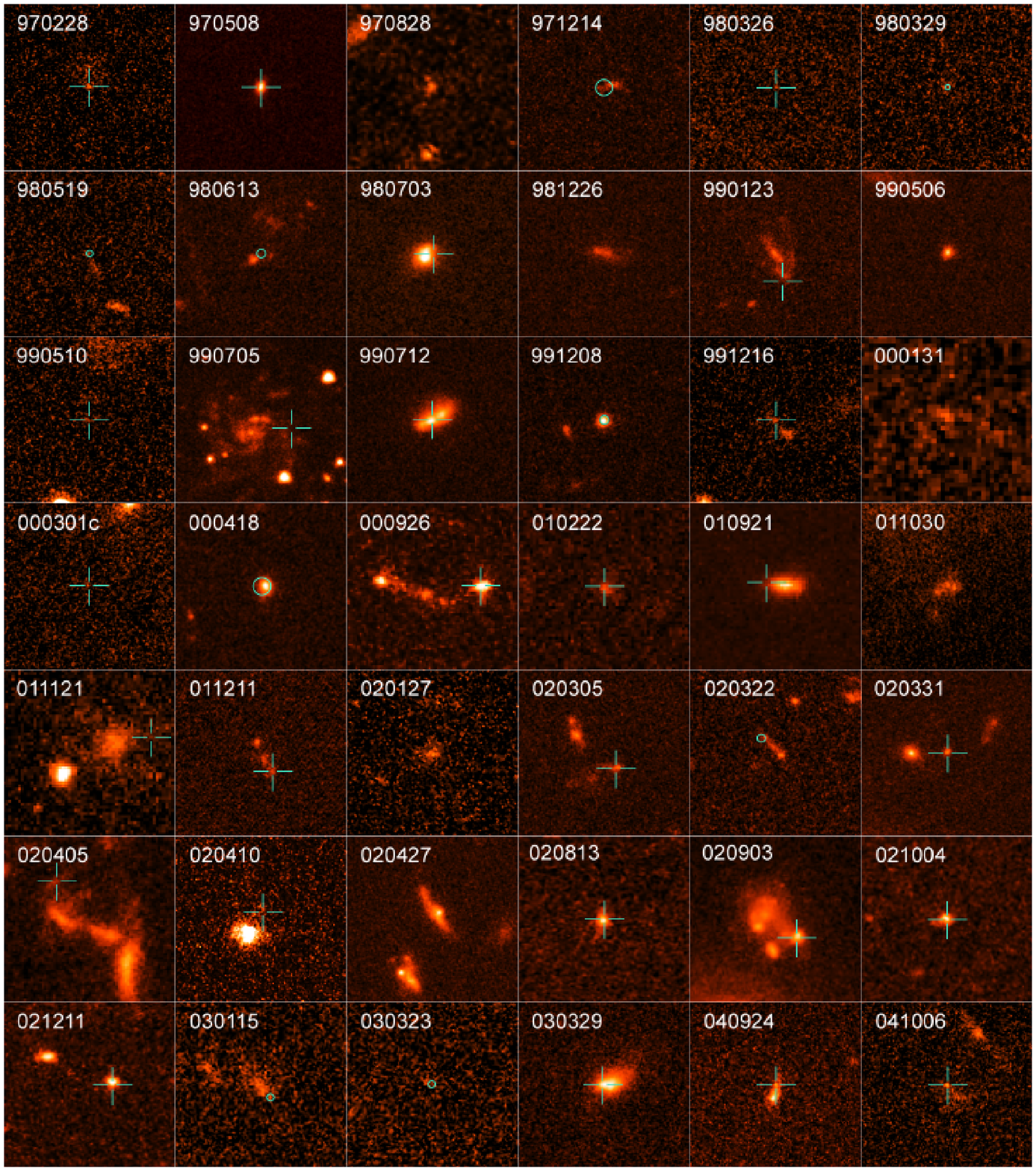,height=6in,angle=0}
\end{center}
\caption{}
\label{fig1}
\end{figure*}

\begin{figure*}[h]
\begin{center}
\psfig{figure=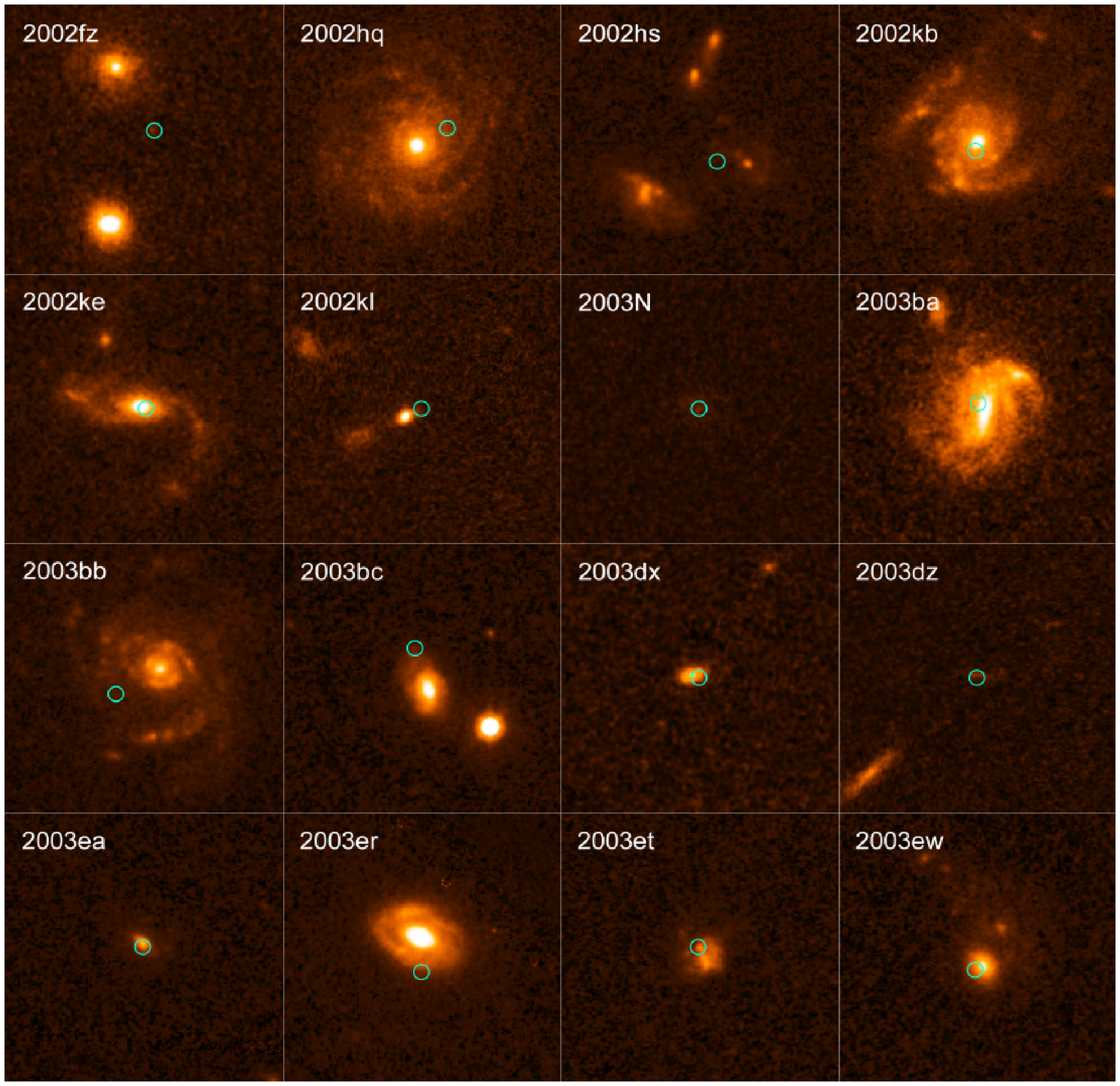,width=15cm,angle=0}
\end{center}
\caption{}
\label{fig1}
\end{figure*}

\begin{figure*}[h]
\begin{center}
\psfig{figure=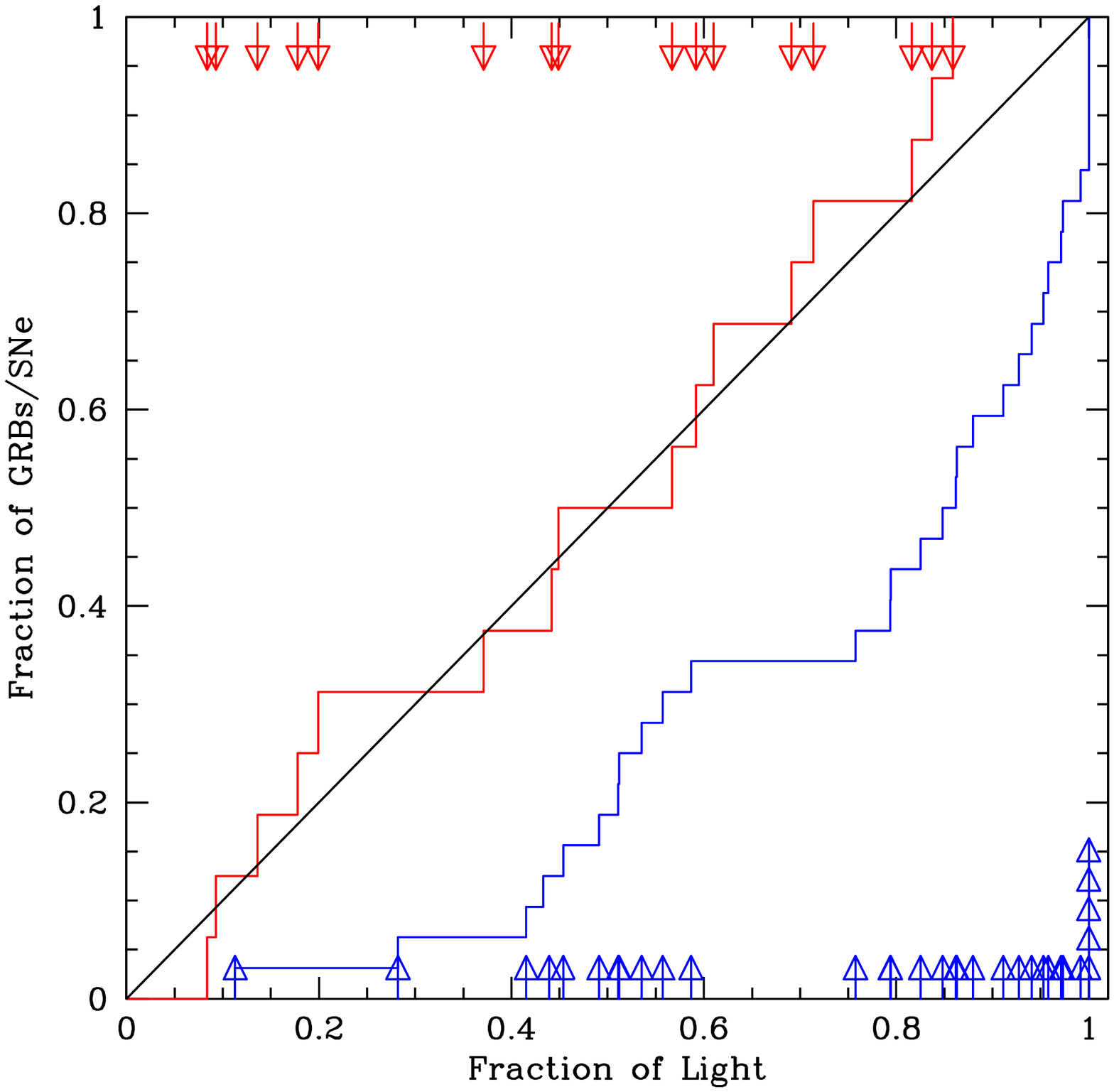,width=5.5in}
\vspace{-1.0cm}
\end{center}
\caption{}   
\end{figure*}

\begin{figure*}[h]
\begin{center}
\psfig{figure=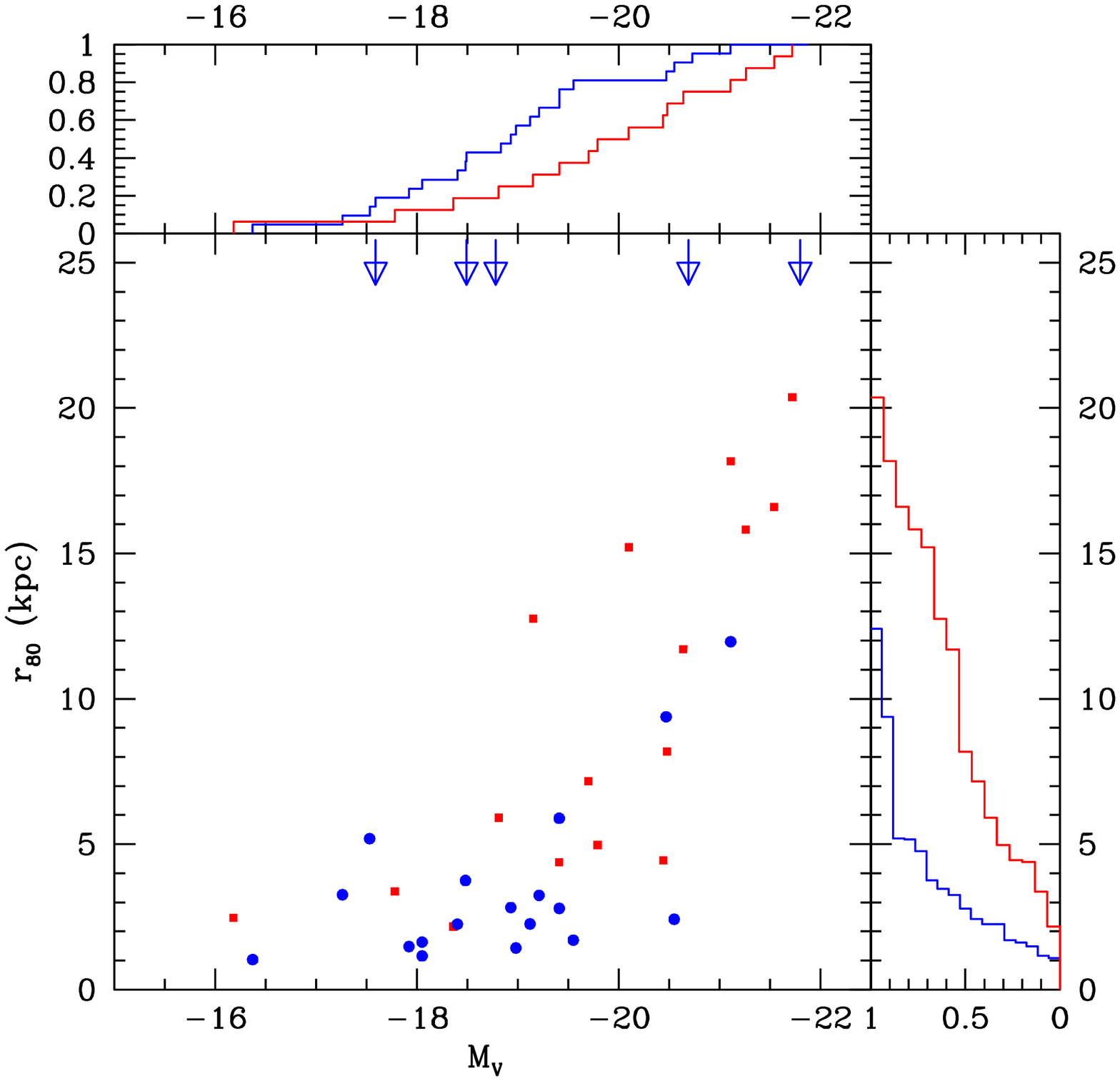,width=5.5in}
\vspace{-1.0cm}
\end{center}
\caption{
}
\end{figure*}

\clearpage


{\bf References}

\begin{thebibliography}{10}
\expandafter\ifx\csname url\endcsname\relax
  \def\url#1{\texttt{#1}}\fi
\expandafter\ifx\csname urlprefix\endcsname\relax\def\urlprefix{URL }\fi
\providecommand{\bibinfo}[2]{#2}
\providecommand{\eprint}[2][]{\url{#2}}

\bibitem{rfc+98}
\bibinfo{author}{{Riess}, A.~G.} \emph{et~al.}
\newblock \bibinfo{title}{{Observational Evidence from Supernovae for an
  Accelerating Universe and a Cosmological Constant}}.
\newblock \emph{\bibinfo{journal}{Astron. J.}} \textbf{\bibinfo{volume}{116}},
  \bibinfo{pages}{1009--1038} (\bibinfo{year}{1998}).

\bibitem{pag+99}
\bibinfo{author}{{Perlmutter}, S.} \emph{et~al.}
\newblock \bibinfo{title}{{Measurements of Omega and Lambda from 42
  High-Redshift Supernovae}}.
\newblock \emph{\bibinfo{journal}{Astrophys. J.}}
  \textbf{\bibinfo{volume}{517}}, \bibinfo{pages}{565--586}
  (\bibinfo{year}{1999}).

\bibitem{bly+05}
\bibinfo{author}{{Branch}, D.}, \bibinfo{author}{{Livio}, M.},
  \bibinfo{author}{{Yungelson}, L.~R.}, \bibinfo{author}{{Boffi}, F.~R.} \&
  \bibinfo{author}{{Baron}, E.}
\newblock \bibinfo{title}{{In Search of the Progenitors of Type IA
  Supernovae}}.
\newblock \emph{\bibinfo{journal}{Publ. Astr. Soc. Pacific}}
  \textbf{\bibinfo{volume}{107}}, \bibinfo{pages}{1019--1028}
  (\bibinfo{year}{1995}).

\bibitem{hfw+03}
\bibinfo{author}{{Heger}, A.}, \bibinfo{author}{{Fryer}, C.~L.},
  \bibinfo{author}{{Woosley}, S.~E.}, \bibinfo{author}{{Langer}, N.} \&
  \bibinfo{author}{{Hartmann}, D.~H.}
\newblock \bibinfo{title}{{How Massive Single Stars End Their Life}}.
\newblock \emph{\bibinfo{journal}{Astrophys. J.}}
  \textbf{\bibinfo{volume}{591}}, \bibinfo{pages}{288--300}
  (\bibinfo{year}{2003}).

\bibitem{kmf+93}
\bibinfo{author}{Kouveliotou, C.} \emph{et~al.}
\newblock \bibinfo{title}{Identification of two classes of gamma-ray bursts}.
\newblock \emph{\bibinfo{journal}{Astrophys. J.}}
  \textbf{\bibinfo{volume}{413}}, \bibinfo{pages}{101--104}
  (\bibinfo{year}{1993}).

\bibitem{gso+05}
\bibinfo{author}{{Gehrels}, N.} \emph{et~al.}
\newblock \bibinfo{title}{{A short {$\gamma$}-ray burst apparently associated
  with an elliptical galaxy at redshift z = 0.225}}.
\newblock \emph{\bibinfo{journal}{Nature}} \textbf{\bibinfo{volume}{437}},
  \bibinfo{pages}{851--854} (\bibinfo{year}{2005}).

\bibitem{pbc+05}
\bibinfo{author}{{Prochaska}, J.~X.} \emph{et~al.}
\newblock \bibinfo{title}{The galaxy hosts and large-scale environments of
  short-hard $\gamma$-ray bursts}.
\newblock \emph{\bibinfo{journal}{Astrophys. J.}}  (\bibinfo{year}{2006}).
\newblock \bibinfo{note}{Accepted, astro-ph/0510022}.

\bibitem{ftm+99}
\bibinfo{author}{Fruchter, A.~S.} \emph{et~al.}
\newblock \bibinfo{title}{{HST} and {Palomar} imaging of {GRB}~990123:
  Implications for the nature of gamma-ray bursts and their hosts}.
\newblock \emph{\bibinfo{journal}{Astrophys. J.}}
  \textbf{\bibinfo{volume}{519}}, \bibinfo{pages}{13--16}
  (\bibinfo{year}{1999}).

\bibitem{sfc+01}
\bibinfo{author}{{Sokolov}, V.~V.} \emph{et~al.}
\newblock \bibinfo{title}{{Host galaxies of gamma-ray bursts: Spectral energy
  distributions and internal extinction}}.
\newblock \emph{\bibinfo{journal}{Astr. Astrophys.}}
  \textbf{\bibinfo{volume}{372}}, \bibinfo{pages}{438--455}
  (\bibinfo{year}{2001}).

\bibitem{ldm+03}
\bibinfo{author}{{Le Floc'h}, E.} \emph{et~al.}
\newblock \bibinfo{title}{{Are the hosts of gamma-ray bursts sub-luminous and
  blue galaxies?}}
\newblock \emph{\bibinfo{journal}{Astr. Astrophys.}}
  \textbf{\bibinfo{volume}{400}}, \bibinfo{pages}{499--510}
  (\bibinfo{year}{2003}).

\bibitem{chg04}
\bibinfo{author}{{Christensen}, L.}, \bibinfo{author}{{Hjorth}, J.} \&
  \bibinfo{author}{{Gorosabel}, J.}
\newblock \bibinfo{title}{{UV star-formation rates of GRB host galaxies}}.
\newblock \emph{\bibinfo{journal}{Astr. Astrophys.}}
  \textbf{\bibinfo{volume}{425}}, \bibinfo{pages}{913--926}
  (\bibinfo{year}{2004}).

\bibitem{bdkf98}
\bibinfo{author}{Bloom, J.~S.}, \bibinfo{author}{Djorgovski, S.~G.},
  \bibinfo{author}{Kulkarni, S.~R.} \& \bibinfo{author}{Frail, D.~A.}
\newblock \bibinfo{title}{The host galaxy of {GRB} 970508}.
\newblock \emph{\bibinfo{journal}{Astrophys. J.}}
  \textbf{\bibinfo{volume}{507}}, \bibinfo{pages}{L25--L28}
  (\bibinfo{year}{1998}).

\bibitem{vfk+01}
\bibinfo{author}{{Vreeswijk}, P.~M.} \emph{et~al.}
\newblock \bibinfo{title}{{VLT} spectroscopy of {GRB} 990510 and {GRB} 990712:
  Probing the faint and bright ends of the gamma-ray burst host galaxy
  population}.
\newblock \emph{\bibinfo{journal}{Astrophys. J.}}
  \textbf{\bibinfo{volume}{546}}, \bibinfo{pages}{672--680}
  (\bibinfo{year}{2001}).

\bibitem{bkd+99}
\bibinfo{author}{{Bloom}, J.~S.} \emph{et~al.}
\newblock \bibinfo{title}{{The unusual afterglow of the gamma-ray burst of 26
  March 1998 as evidence for a supernova connection.}}
\newblock \emph{\bibinfo{journal}{Nature}} \textbf{\bibinfo{volume}{401}},
  \bibinfo{pages}{453--456} (\bibinfo{year}{1999}).

\bibitem{gtv+00}
\bibinfo{author}{{Galama}, T.~J.} \emph{et~al.}
\newblock \bibinfo{title}{Evidence for a supernova in reanalyzed optical and
  near-infrared images of {GRB} 970228}.
\newblock \emph{\bibinfo{journal}{Astrophys. J.}}
  \textbf{\bibinfo{volume}{536}}, \bibinfo{pages}{185--194}
  (\bibinfo{year}{2000}).

\bibitem{lnf+05}
\bibinfo{author}{{Levan}, A.} \emph{et~al.}
\newblock \bibinfo{title}{{GRB 020410: A Gamma-Ray Burst Afterglow Discovered
  by Its Supernova Light}}.
\newblock \emph{\bibinfo{journal}{Astrophys. J.}}
  \textbf{\bibinfo{volume}{624}}, \bibinfo{pages}{880--888}
  (\bibinfo{year}{2005}).

\bibitem{hsm+03}
\bibinfo{author}{{Hjorth}, J.} \emph{et~al.}
\newblock \bibinfo{title}{{A very energetic supernova associated with the
  {$\gamma$}-ray burst of 29 March 2003}}.
\newblock \emph{\bibinfo{journal}{Nature}} \textbf{\bibinfo{volume}{423}},
  \bibinfo{pages}{847--850} (\bibinfo{year}{2003}).

\bibitem{smg+03}
\bibinfo{author}{{Stanek}, K.~Z.} \emph{et~al.}
\newblock \bibinfo{title}{{Spectroscopic Discovery of the Supernova 2003dh
  Associated with GRB\,030329}}.
\newblock \emph{\bibinfo{journal}{Astrophys. J.}}
  \textbf{\bibinfo{volume}{591}}, \bibinfo{pages}{L17--L20}
  (\bibinfo{year}{2003}).

\bibitem{dmb+03}
\bibinfo{author}{{Della Valle}, M.} \emph{et~al.}
\newblock \bibinfo{title}{{Evidence for supernova signatures in the spectrum of
  the late-time bump of the optical afterglow of GRB 021211}}.
\newblock \emph{\bibinfo{journal}{Astr. Astrophys.}}
  \textbf{\bibinfo{volume}{406}}, \bibinfo{pages}{L33--L37}
  (\bibinfo{year}{2003}).

\bibitem{mtc+04}
\bibinfo{author}{{Malesani}, D.} \emph{et~al.}
\newblock \bibinfo{title}{{SN 2003lw and GRB 031203: A Bright Supernova for a
  Faint Gamma-Ray Burst}}.
\newblock \emph{\bibinfo{journal}{Astrophys. J.}}
  \textbf{\bibinfo{volume}{609}}, \bibinfo{pages}{L5--L8}
  (\bibinfo{year}{2004}).

\bibitem{zkh04}
\bibinfo{author}{{Zeh}, A.}, \bibinfo{author}{{Klose}, S.} \&
  \bibinfo{author}{{Hartmann}, D.~H.}
\newblock \bibinfo{title}{{A Systematic Analysis of Supernova Light in
  Gamma-Ray Burst Afterglows}}.
\newblock \emph{\bibinfo{journal}{Astrophys. J.}}
  \textbf{\bibinfo{volume}{609}}, \bibinfo{pages}{952--961}
  (\bibinfo{year}{2004}).

\bibitem{pk01}
\bibinfo{author}{{Panaitescu}, A.} \& \bibinfo{author}{{Kumar}, P.}
\newblock \bibinfo{title}{{Fundamental Physical Parameters of Collimated
  Gamma-Ray Burst Afterglows}}.
\newblock \emph{\bibinfo{journal}{Astrophys. J.}}
  \textbf{\bibinfo{volume}{560}}, \bibinfo{pages}{L49--L53}
  (\bibinfo{year}{2001}).

\bibitem{fks+01}
\bibinfo{author}{{Frail}, D.~A.} \emph{et~al.}
\newblock \bibinfo{title}{{Beaming in Gamma-Ray Bursts: Evidence for a Standard
  Energy Reservoir}}.
\newblock \emph{\bibinfo{journal}{Astrophys. J.}}
  \textbf{\bibinfo{volume}{562}}, \bibinfo{pages}{L55--58}
  (\bibinfo{year}{2001}).

\bibitem{mwh01}
\bibinfo{author}{{MacFadyen}, A.~I.}, \bibinfo{author}{{Woosley}, S.~E.} \&
  \bibinfo{author}{{Heger}, A.}
\newblock \bibinfo{title}{{Supernovae, Jets, and Collapsars}}.
\newblock \emph{\bibinfo{journal}{Astrophys. J.}}
  \textbf{\bibinfo{volume}{550}}, \bibinfo{pages}{410--425}
  (\bibinfo{year}{2001}).

\bibitem{vhf96}
\bibinfo{author}{{van Dyk}, S.~D.}, \bibinfo{author}{{Hamuy}, M.} \&
  \bibinfo{author}{{Filippenko}, A.~V.}
\newblock \bibinfo{title}{{Supernovae and Massive Star Formation Regions}}.
\newblock \emph{\bibinfo{journal}{Astron. J.}} \textbf{\bibinfo{volume}{111}},
  \bibinfo{pages}{2017--2027} (\bibinfo{year}{1996}).

\bibitem{vdblf05}
\bibinfo{author}{{van den Bergh}, S.}, \bibinfo{author}{{Li}, W.} \&
  \bibinfo{author}{{Filippenko}, A.~V.}
\newblock \bibinfo{title}{{Classifications of the Host Galaxies of Supernovae,
  Set III}}.
\newblock \emph{\bibinfo{journal}{Publ. Astr. Soc. Pacific}}
  \textbf{\bibinfo{volume}{117}}, \bibinfo{pages}{773--782}
  (\bibinfo{year}{2005}).

\bibitem{pmn+04}
\bibinfo{author}{{Podsiadlowski}, P.}, \bibinfo{author}{{Mazzali}, P.~A.},
  \bibinfo{author}{{Nomoto}, K.}, \bibinfo{author}{{Lazzati}, D.} \&
  \bibinfo{author}{{Cappellaro}, E.}
\newblock \bibinfo{title}{{The Rates of Hypernovae and Gamma-Ray Bursts:
  Implications for Their Progenitors}}.
\newblock \emph{\bibinfo{journal}{Astrophys. J.}}
  \textbf{\bibinfo{volume}{607}}, \bibinfo{pages}{L17--L20}
  (\bibinfo{year}{2004}).

\bibitem{bkd02}
\bibinfo{author}{{Bloom}, J.~S.}, \bibinfo{author}{{Kulkarni}, S.~R.} \&
  \bibinfo{author}{{Djorgovski}, S.~G.}
\newblock \bibinfo{title}{{The Observed Offset Distribution of Gamma-Ray Bursts
  from Their Host Galaxies: A Robust Clue to the Nature of the Progenitors}}.
\newblock \emph{\bibinfo{journal}{Astron. J.}} \textbf{\bibinfo{volume}{123}},
  \bibinfo{pages}{1111--1148} (\bibinfo{year}{2002}).

\bibitem{rst+04}
\bibinfo{author}{{Riess}, A.~G.} \emph{et~al.}
\newblock \bibinfo{title}{{Identification of Type Ia Supernovae at Redshift 1.3
  and Beyond with the Advanced Camera for Surveys on the Hubble Space
  Telescope}}.
\newblock \emph{\bibinfo{journal}{Astrophys. J.}}
  \textbf{\bibinfo{volume}{600}}, \bibinfo{pages}{L163--L166}
  (\bibinfo{year}{2004}).

\bibitem{srd+04}
\bibinfo{author}{{Strolger}, L.-G.} \emph{et~al.}
\newblock \bibinfo{title}{{The Hubble Higher z Supernova Search: Supernovae to
  z \~{} 1.6 and Constraints on Type Ia Progenitor Models}}.
\newblock \emph{\bibinfo{journal}{Astrophys. J.}}
  \textbf{\bibinfo{volume}{613}}, \bibinfo{pages}{200--223}
  (\bibinfo{year}{2004}).

\bibitem{gfk+04}
\bibinfo{author}{{Giavalisco}, M.} \emph{et~al.}
\newblock \bibinfo{title}{{The Great Observatories Origins Deep Survey: Initial
  Results from Optical and Near-Infrared Imaging}}.
\newblock \emph{\bibinfo{journal}{Astrophys. J.}}
  \textbf{\bibinfo{volume}{600}}, \bibinfo{pages}{L93--L98}
  (\bibinfo{year}{2004}).

\bibitem{cvf+05}
\bibinfo{author}{{Conselice}, C.~J.} \emph{et~al.}
\newblock \bibinfo{title}{{Gamma-Ray Burst Selected High Redshift Galaxies:
  Comparison to Field Galaxy Populations to z\~{}3}}.
\newblock \emph{\bibinfo{journal}{Astrophys. J.}}
  \textbf{\bibinfo{volume}{633}}, \bibinfo{pages}{29--40}
  (\bibinfo{year}{2005}).

\bibitem{sfd98}
\bibinfo{author}{Schlegel, D.~J.}, \bibinfo{author}{Finkbeiner, D.~P.} \&
  \bibinfo{author}{Davis, M.}
\newblock \bibinfo{title}{Maps of dust infrared emission for use in estimation
  of reddening and cosmic microwave background radiation foregrounds}.
\newblock \emph{\bibinfo{journal}{Astrophys. J.}}
  \textbf{\bibinfo{volume}{500}}, \bibinfo{pages}{525--553}
  (\bibinfo{year}{1998}).

\bibitem{mhc+03}
\bibinfo{author}{{Mirabal}, N.} \emph{et~al.}
\newblock \bibinfo{title}{{GRB 021004: A Possible Shell Nebula around a
  Wolf-Rayet Star Gamma-Ray Burst Progenitor}}.
\newblock \emph{\bibinfo{journal}{Astrophys. J.}}
  \textbf{\bibinfo{volume}{595}}, \bibinfo{pages}{935--949}
  (\bibinfo{year}{2003}).

\bibitem{kgr+04}
\bibinfo{author}{{Klose}, S.} \emph{et~al.}
\newblock \bibinfo{title}{{Probing a Gamma-Ray Burst Progenitor at a Redshift
  of z = 2: A Comprehensive Observing Campaign of the Afterglow of GRB
  030226}}.
\newblock \emph{\bibinfo{journal}{Astron. J.}} \textbf{\bibinfo{volume}{128}},
  \bibinfo{pages}{1942--1954} (\bibinfo{year}{2004}).

\bibitem{wk05b}
\bibinfo{author}{Weidner, C.} \& \bibinfo{author}{Kroupa, P.}
\newblock \bibinfo{title}{Variations of the imf}.
\newblock In \bibinfo{editor}{Corbelli, E.}, \bibinfo{editor}{Plla, F.} \&
  \bibinfo{editor}{Zinnecker, H.} (eds.) \emph{\bibinfo{booktitle}{The Initial
  Mass Function 50 Year Later}}, \bibinfo{pages}{125--186}
  (\bibinfo{publisher}{Springer}, \bibinfo{address}{Dordrecht, The
  Netherlands}, \bibinfo{year}{2005}).

\bibitem{bfs+05}
\bibinfo{author}{Bersier, D.} \emph{et~al.}
\newblock \bibinfo{title}{Evidence for a supernova associated with the x-ray
  flash 020903}.
\newblock \emph{\bibinfo{journal}{Astrophys. J.}}  (\bibinfo{year}{2005}).
\newblock \bibinfo{note}{Submitted to Ap. J.}

\bibitem{vel+04}
\bibinfo{author}{{Vreeswijk}, P.~M.} \emph{et~al.}
\newblock \bibinfo{title}{{The host of GRB 030323 at z=3.372: A very high
  column density DLA system with a low metallicity}}.
\newblock \emph{\bibinfo{journal}{Astr. Astrophys.}}
  \textbf{\bibinfo{volume}{419}}, \bibinfo{pages}{927--940}
  (\bibinfo{year}{2004}).

\bibitem{pbc+04}
\bibinfo{author}{{Prochaska}, J.~X.} \emph{et~al.}
\newblock \bibinfo{title}{{The Host Galaxy of GRB 031203: Implications of Its
  Low Metallicity, Low Redshift, and Starburst Nature}}.
\newblock \emph{\bibinfo{journal}{Astrophys. J.}}
  \textbf{\bibinfo{volume}{611}}, \bibinfo{pages}{200--207}
  (\bibinfo{year}{2004}).

\bibitem{cpb+05}
\bibinfo{author}{{Chen}, H.-W.}, \bibinfo{author}{{Prochaska}, J.~X.},
  \bibinfo{author}{{Bloom}, J.~S.} \& \bibinfo{author}{{Thompson}, I.~B.}
\newblock \bibinfo{title}{{Echelle Spectroscopy of a GRB Afterglow at z=3.969:
  A New Probe of the Interstellar and Intergalactic Media in the Young
  Universe}}.
\newblock \emph{\bibinfo{journal}{Astrophys. J.}}
  \textbf{\bibinfo{volume}{634}}, \bibinfo{pages}{L25--L28}
  (\bibinfo{year}{2005}).

\bibitem{kk04}
\bibinfo{author}{{Kobulnicky}, H.~A.} \& \bibinfo{author}{{Kewley}, L.~J.}
\newblock \bibinfo{title}{Metallicities of galaxies in the {GOODS-North
  Field}}.
\newblock \emph{\bibinfo{journal}{Astrophys. J.}}
  \textbf{\bibinfo{volume}{617}}, \bibinfo{pages}{240--261}
  (\bibinfo{year}{2004}).

\bibitem{fng+05}
\bibinfo{author}{{Figer}, D.~F.}, \bibinfo{author}{{Najarro}, F.},
  \bibinfo{author}{{Geballe}, T.~R.}, \bibinfo{author}{{Blum}, R.~D.} \&
  \bibinfo{author}{{Kudritzki}, R.~P.}
\newblock \bibinfo{title}{{Massive Stars in the SGR 1806-20 Cluster}}.
\newblock \emph{\bibinfo{journal}{Astrophys. J.}}
  \textbf{\bibinfo{volume}{622}}, \bibinfo{pages}{L49--L52}
  (\bibinfo{year}{2005}).

\bibitem{ch06}
\bibinfo{author}{Crowther, P.~A.} \& \bibinfo{author}{Hadfield, L.~J.}
\newblock \bibinfo{title}{Reduced wolf-rayet line luminosities at low
  metallicity}.
\newblock \emph{\bibinfo{journal}{Astr. Astrophys.}} \bibinfo{pages}{711--722}
  (\bibinfo{year}{2006}).

\bibitem{wh05}
\bibinfo{author}{{Woosley}, S.} \& \bibinfo{author}{{Heger}, A.}
\newblock \bibinfo{title}{{The Progenitor Stars of Gamma-Ray Bursts}}.
\newblock \emph{\bibinfo{journal}{Astrophys. J.}}
  \textbf{\bibinfo{volume}{637}}, \bibinfo{pages}{914--921}
  (\bibinfo{year}{2006}).

\bibitem{yl05}
\bibinfo{author}{{Yoon}, S.-C.} \& \bibinfo{author}{{Langer}, N.}
\newblock \bibinfo{title}{{Evolution of rapidly rotating metal-poor massive
  stars towards gamma-ray bursts}}.
\newblock \emph{\bibinfo{journal}{Astr. Astrophys.}} \bibinfo{pages}{643--648}
  (\bibinfo{year}{2005}).

\bibitem{cbs+05}
\bibinfo{author}{{Chapman}, S.~C.}, \bibinfo{author}{{Blain}, A.~W.},
  \bibinfo{author}{{Smail}, I.} \& \bibinfo{author}{{Ivison}, R.~J.}
\newblock \bibinfo{title}{{A Redshift Survey of the Submillimeter Galaxy
  Population}}.
\newblock \emph{\bibinfo{journal}{Astrophys. J.}}
  \textbf{\bibinfo{volume}{622}}, \bibinfo{pages}{772--796}
  (\bibinfo{year}{2005}).

\bibitem{fjm+03}
\bibinfo{author}{{Fynbo}, J.~P.~U.} \emph{et~al.}
\newblock \bibinfo{title}{{On the Lyalpha emission from gamma-ray burst host
  galaxies: Evidence for low metallicities}}.
\newblock \emph{\bibinfo{journal}{Astr. Astrophys.}}
  \textbf{\bibinfo{volume}{406}}, \bibinfo{pages}{L63--L66}
  (\bibinfo{year}{2003}).

\bibitem{snk+06}
\bibinfo{author}{{Soderberg}, A.~M.}, \bibinfo{author}{{Nakar}, E.},
  \bibinfo{author}{{Kulkarni}, S.~R.} \& \bibinfo{author}{{Berger}, E.}
\newblock \bibinfo{title}{{Late-time Radio Observations of 68 Type Ibc
  Supernovae: Strong Constraints on Off-Axis Gamma-ray Bursts}}.
\newblock \emph{\bibinfo{journal}{Astrophys. J.}}
  \textbf{\bibinfo{volume}{638}}, \bibinfo{pages}{930--937}
  (\bibinfo{year}{2006}).

\bibitem{vdbt91}
\bibinfo{author}{{van den Bergh}, S.} \& \bibinfo{author}{{Tammann}, G.~A.}
\newblock \bibinfo{title}{{Galactic and extragalactic supernova rates}}.
\newblock \emph{\bibinfo{journal}{Ann. Rev. Astr. Ap.}}
  \textbf{\bibinfo{volume}{29}}, \bibinfo{pages}{363--407}
  (\bibinfo{year}{1991}).

\bibitem{mdvp+05}
\bibinfo{author}{{Mannucci}, F.} \emph{et~al.}
\newblock \bibinfo{title}{{The supernova rate per unit mass}}.
\newblock \emph{\bibinfo{journal}{Astr. Astrophys.}}
  \textbf{\bibinfo{volume}{433}}, \bibinfo{pages}{807--814}
  (\bibinfo{year}{2005}).

\end{thebibliography}
\end{document}